\RequirePackage{ifpdf}
\ifpdf 
\documentclass[pdftex]{sigma}
\else
\documentclass{sigma}
\fi

\begin{document}

\renewcommand{\PaperNumber}{***}

\FirstPageHeading

\ShortArticleName{Generalized Theta Functions}

\ArticleName{Generalized Theta Functions. I}

\Author{Yuriy P. Smilyanets }

\AuthorNameForHeading{Yuriy P. Smilyanets}

\EmailD{\href{mailto:yusmill@ukr.net}{yusmill@ukr.net}} 




\Abstract{Generalizations of classical theta functions are proposed that include any even number of analytic parameters for which conditions of quasi-periodicity are fulfilled and that are representations of extended Heisenberg group. Differential equations for generalized theta functions and  finite 
 non-unitary representations of extended Heisenberg group are presented so as other properties and possible applications are pointed out such as a projective embedding of tori by means of generalized theta functions.}


\Classification{14K25; 22E70; 81R05}

\section{Introduction}

The work fulfills generalization of classical theta functions (CTF) as they are described following to lectures~\cite{Mumford}. The aim was to use the next terms on summation index in the exponent function of CTF including any even number of analytic parameters for which conditions of quasi-periodicity hold true.
The work was inspired by papers \cite{Scott}, \cite{Janik}  where "nucleon structure functions are shown to have a relation
to the classical elliptic theta functions" and  in \cite{Scott} it was stated that theta functions may be generalized,
and by the paper \cite{Vafa} where it was shown that CTF is the partition function for Hamiltonian of the string theory.

Saving notations for the CTF the formula of the minimal generalization of the CTF with the next two and others up to \(N = 2m\)  not pointed out here for simplicity parameters has the view (see section~\ref{sec: many parameter} for unified notations for general case of \(N = 2m\) number of parameters):
\begin{equation}
\Theta\left(z,\tau,\rho,\delta,\cdots\right)=\
\sum_{n\in Z}exp\lbrace2\pi i n z + \pi i {n^2}\tau +\frac{1}{3}\pi i {n^3}\rho + \frac{1}{12}\pi i{n^4} \delta +\cdots \rbrace,\label{eq1}
\end{equation}
where \(z, \tau, \rho \in{C}\) and \(Im\delta>\)0 in the case of four parameters (for the last parameter in the common case of any even number of them).   
Statements about domains of convergence of generalized theta functions (GTF) 
 are presented in Appendix.

This formula may be written in more short and useful form for any even number of analytic variables:
\begin{gather}
\Theta\left(z,\tau,\rho,\delta,\cdots\right)=\
\sum_{n\in Z}exp\lbrace2\pi i \varphi(n;z,\tau,\rho,\delta,\cdots) \rbrace, \\
\varphi(n) \equiv  \varphi(n; z,\tau,\rho,\delta,\cdots) =  z n + \frac{n^2}{2!} \tau + \frac{n^3}{3!} \rho + \frac{n^4}{4!}  \delta + \cdots =\nonumber\\
= \varphi(n; \varphi^{'}_n(0),\varphi^{''}_{nn}(0),\varphi^{'''}_{nnn}(0),\varphi^{''''}_{nnnn},\cdots).\nonumber
\end{gather}

We see that \( \varphi(n) \) is self determined by its derivatives in zero point of summation parameter \("n"\) analytic function 
that defines, as is shown below in Section \ref{Quasi-periodic properties}, quasi periodic property that is an essence of theta function: to be "unique" per current opinion \cite{Mumford} self determined quasi periodic function of multiple parameters and be representations of generalized Heisenberg group, finite non-unitary representations of which are presented in Section \ref{Representations}. Introducing a pair of new parameters, it provides analytic continuation on parameter \( \tau\) of the classical TF into lower part of complex plane \(C\) and at the same time represents successive and natural hierarchy 
that is so important in physics. 

Generalized theta functions with rational characteristics and projective embedding of tori by means of GTFs with rational characteristics are presented in 
Section \ref{sec: GTF with char}.

Equations for the generalized theta functions are represented in Section \ref{sec: Equations}  the simplest of which and more well-known are  heat-type equations. 
In these processes  imaginary parts of parameters \(\tau\)  and \(\delta\) have a \(\textit{time}\)  character 
relatively to a \(\textit{space}\)  character of parameters \(z\) and \(\tau\)  respectively.
In Section \ref{sec: many parameter} the formulas for unified view in number of analytic parameters are represented so as  relations between parameters of quasi-periodicity are pointed out.

\section{ Quasi-periodic properties of GTF and extended Heisenberg group }
\label{Quasi-periodic properties}

This formula preserves quasi-periodic property that in the case of four singled out parameters has the view:
\begin{gather}
T_a \Theta\left(z,\tau,\rho,\delta,\cdots\right) =\nonumber\\
=\Theta\left(z+a\tau+\frac{a^2}{2}\rho+\frac{a^3}{6}\delta +\cdots,\tau+a\rho+\frac{a^2}{2}\delta+\cdots,\rho+a\delta+\cdots,\delta+\cdots\right)=\nonumber\\
= exp\lbrace -2\pi i za - \pi i{a^2}\tau  - \pi i \frac{a^3}{3} \rho -\pi i \frac{a^4}{12}  \delta -\cdots \rbrace \Theta_a\left(z,\tau,\rho,\delta,\cdots\right),\label{eq2}
\end{gather}
where  \(\Theta_a\left(z,\tau,\rho,\delta,\cdots\right)\) and this formula can be written more concisely as following:
\begin{gather}
T_a \Theta\left(z,\tau,\rho,\delta,\cdots\right) =\Theta\left(\varphi^{'}(a),\varphi^{''}(a),\varphi^{'''}(a),\varphi^{''''}(a),\cdots\right)=\nonumber\\
 = exp\lbrace -2\pi i \varphi(a)\rbrace \Theta\left(\varphi(n+a;z,\tau,\rho,\delta,\cdots)\right) \
  \equiv  exp\lbrace -2\pi i \varphi(a)\rbrace \Theta_a\left(z,\tau,\rho,\delta,\cdots\right),\label{eq3}
\end{gather}
where \( \varphi(a) = \varphi(n=a) =\varphi(a;z,\tau,\rho,\delta,\cdots) =  za + \frac{a^2}{2!} \tau + \frac{a^3}{3!} \rho + \frac{a^4}{4!}  \delta +\cdots \)  and \(\varphi^{'}(a)\) is a derivative on parameter \(a\), or \(\varphi^{'}(a) =\varphi^{'}_a=\varphi^{'}_a(a) = \varphi^{'}_n(n)_{|n=a}\).

Remark, that  \(\varphi^{'}_a(0) = z\),  \(\varphi^{''}_{aa}(0) = \tau\), \(\varphi^{'''}_{aaa}(0) = \rho\), \(\varphi^{''''}_{aaaa}(0) = \delta,\cdots\)  and
equation  \eqref{eq3} means that  in the case of any even number of parameters \(\varphi(n;z,\tau,\rho,\delta,\cdots)\) satisfies equation:
\begin{gather*}
T_a \sum_{n\in Z}exp\lbrace2\pi i  \varphi(n;z,\tau,\rho,\delta,\cdots)\rbrace = \sum_{n\in Z}exp\lbrace2\pi i\varphi(n; \varphi^{'}(a),\varphi^{''}(a),\varphi^{'''}(a),\varphi^{''''}(a),\cdots)\rbrace =\\
= \sum_{n\in Z}exp\lbrace2\pi i[\varphi(n+a;z,\tau,\rho,\delta,\cdots) - \varphi(a;z,\tau,\rho,\delta,\cdots)]\rbrace,\\
\varphi(n; \varphi^{'}(a),\varphi^{''}(a),\varphi^{'''}(a),\varphi^{''''}(a),\cdots) = \varphi(n+a;z,\tau,\rho,\delta,\cdots) - \varphi(a;z,\tau,\rho,\delta,\cdots) =\\
=n \varphi^{'}(a)+\frac{n^2}{2!}\varphi^{''}(a) + \frac{n^3}{3!}\varphi^{'''}(a)  + \frac{n^4}{4!}\varphi^{''''}(a) + \cdots.
\end{gather*}


Equations \eqref{eq2} and \eqref{eq3} for operators \(T_{a1}\) and \(T_{a2}\) satisfy relation: \(   T_{a1}T_{a2} = T_{a1+a2}\)
 that together with equations for translation operator \(S_b\) and commutator of  \(S_b\) and  \(T_a\) 
compose well known  Heisenberg group (HG) \cite{Mumford}:
\begin{gather}
T_{a1}T_{a2} = T_{a1+a2}, \nonumber\\
 S_b \Theta\left(z,\tau,\rho,\delta,\cdots\right) = \Theta\left(z +  b,\tau,\rho,\delta,\cdots\right),\\     
  T_a S_b  =exp\lbrace -2\pi i ab \rbrace  S_b T_a. \nonumber
\end{gather}

We see that there exist many parameter representations of the  Heisenberg group.

Besides these operators there are others:
\begin{gather}
S_{b,c,d,e,\cdots}\Theta\left(z ,\tau,\rho,\delta,\cdots\right) = \Theta\left(z + b,\tau + c,\rho +d,\delta + e,\cdots\right),\\
T_a S_{b,c,d,e,\cdots}  = exp\lbrace -2\pi i ab -  \pi i{a^2}c  - \pi i \frac{a^3}{3}d -\pi i \frac{a^4}{12} e -\cdots\rbrace S_{b,c,d,e,\cdots} T_a =\nonumber\\
= exp\lbrace -2\pi i \varphi(a;b,c,d,e,\cdots) \rbrace S_{b,c,d,e,\cdots} T_a.\label{eq7}
\end{gather}
\(\Theta\left(z ,\tau,\rho,\delta,\cdots\right)\)  function is invariant under translational subgroup \( S_{b,2c,6d,24e,\cdots}\), where 

\(b,c,d,e,\cdots\in Z\).

The action of the element (\(\lambda,a,b,c,d,e,\cdots)\) of the extended Heisenberg group (EHG)
 of quasi periodic transformations, where \(\lambda \) denotes multiplication on complex number with unit norm, may be written as in \cite{Mumford}: 
\begin{gather}
(\lambda,a,b,c,d,e,\cdots)(\lambda',a',b',c',d',e',\cdots) =\nonumber\\
=(\lambda\lambda'exp\lbrace -2\pi i [ab' -  \frac{a^2}{2!}c'  - \frac{a^3}{3!}d' -\frac{a^4}{4!} e',\cdots]\rbrace ,a + a' ,b+b',c+c',d+d',e+e',\cdots) =\nonumber\\
=(\lambda\lambda'exp\lbrace -2\pi i \varphi(a;b',c',d',e',\cdots)\rbrace,a+a',b+b',c+c',d+d',e+e',\cdots).
\end{gather}
So \(\Theta\left(z ,\tau,\rho,\delta,\cdots\right)\)  function is representation of the extended Heisenberg group (\(\lambda,a,b,c,d,e,...\)) and the view of equation \eqref{eq3} gives an extended sight on the true nature of the theta functions.

\section{Representations of the EHG in the space of parameters }
\label{Representations}

Omitting in this section for simplicity "three dots",we can made a {\bfseries Statement 3.1:} {\itshape The action of \(\lambda S_{b,c,d,e} T_a\) elements of the extended Heisenberg group  on parameters can be expressed by the equation for non-unitary finite matrix representation:}
\begin{gather*}
\left(\begin{array}{cccccc}
1&a&a^2/2!&a^3/3!&a^4/4!&-\frac{ln\lambda_1}{2\pi i}\\
0&1&a&a^2/2!&a^3/3!&b\\
0&0&1&a&a^2/2!&c\\
0&0&0&1&a&d\\
0&0&0&0&1&e\\
0&0&0&0&0&1
\end{array}\right) 
\left(\begin{array}{c}
-\frac{ln\lambda_2}{2\pi i}\\
z\\
\tau\\
\rho\\
\delta\\
1
\end{array}\right)
=\\
=
\left(\begin{array}{c}
-\frac{ln(\lambda_1\lambda_2)}{2\pi i} + az + (a^2/2!)\tau + (a^3/3!)\rho + (a^4/4!)\delta\\ 
z + b + a\tau + (a^2/2!)\rho + (a^3/3!)\delta\\
\tau + c + a\rho + (a^2/2!)\delta\\
\rho + d + a\delta\\
\delta + e\\
1
\end{array}\right),
\end{gather*}
{\itshape that may be regarded as solution of equation 
\begin{equation*}
\hat A_a(-\frac{ln\lambda_2}{2\pi i} ,z,\tau,\rho,\delta,1)^t = (-\frac{ln(\lambda_1\lambda_2)}{2\pi i} + \varphi(a),\varphi^{'}(a),\varphi^{''}(a),\varphi^{'''}(a),\varphi^{''''}(a),1)^t 
\end{equation*}
 relatively to matrix operator \(\hat A_a\) for a given function \(\varphi(a) = \varphi(a,z,\tau,\rho,\delta)\) 
that evidently satisfies the relation \(\hat A_{a2} \hat A_{a1} = \hat A_{a1+a2}\) the same as \(T_{a2} T_{a1} = T_{a1+a2}\) and other EHG equations and may be marked as \(EH_6(R)\) that is different from the standard \(H_3(R)\) finite non-unitary representations of the HG that is in the right low corner \(3\times 3\) part of the matrix.}

Many parameter non-unitary representation may be represented in the form similar to the infinite dimensional one 
by the matrix equation:
\begin{equation*}
\left(\begin{array}{ccccccc}
1&0&0&0&0&0&0\\
-\frac{ln\lambda}{2\pi i}&1&a&a^2/2&a^3/6&a^4/24&...\\
b&0&1&a&a^2/2&a^3/6&...\\
c&0&0&1&a&a^2/2&...\\
d&0&0&0&1&a&...\\
e&0&0&0&0&1&...\\
\vdots&\vdots&\vdots&\vdots&\vdots&\vdots&\vdots\vdots\vdots
\end{array}\right) 
\left(\begin{array}{c}
1\\
0\\
z\\
\tau\\
\rho\\
\delta\\
\vdots
\end{array}\right)
=
\left(\begin{array}{c}
1\\
-\frac{ln\lambda}{2\pi i} + az + (a^2/2)\tau + (a^3/6)\rho + (a^4/24)\delta\\ 
z + b + a\tau + (a^2/2)\rho + (a^3/6)\delta\cdots\\
\tau + c + a\rho + (a^2/2)\delta\cdots\\
\rho + d + a\delta\cdots\\
\delta + e\cdots\\
\vdots
\end{array}\right),
\end{equation*}

but not equal to taking into account pointed out conditions of convergence.

\section[Theta functions with characteristics]{Generalized theta functions with characteristics }
\label{sec: GTF with char}

Generalized theta functions 
with rational characteristics 
 are obtained by transformations
\begin{equation*}
 \Theta  \Big [^{~~a~}_{b,c,d,...} \Big ]\left(z ,\tau,\rho,\delta,\cdots\right) =  S_{b,c,d,...} T_a \Theta\left(z ,\tau,\rho,\delta,\cdots\right), \nonumber\\
\end{equation*}
or in the obvious form:
\begin{equation*}
 \Theta \Big [ ^{~~a~}_{b,c,d,...} \Big ]\left(z ,\tau,\rho,\delta,\cdots\right) = \\
 \sum_{n\in Z}exp\lbrace2\pi i [(n + a) (z + b) + \frac{1}{2!}{(n + a)^2}(\tau + c) +\frac{1}{3!}{(n + a)^3}(\rho + d) + \cdots] \rbrace.
\end{equation*}

Relations between elements of the extended Heisenberg group denote relations and properties of the generalized theta functions with characteristics.

In the case of four variables GTFs with characteristics  are defined as 
 \(\Theta \big [ ^{~~a~}_{b,2!c,3!d} \big ](z,\tau,\rho,\delta)\), where~ a,d~\(\in((1/l) Z/Z)\), 
\hspace{1mm} c~\(\in((1/l^{2}) Z/Z)\),\hspace{1mm} b~\(\in((1/l^{3}) Z/Z)\)
of common dimension \(l^{7} \) .

Introduce complex torus  \( E_{\tau,\rho,\delta,\cdots} = C/\Lambda_{\tau,\rho,\delta,\cdots} \),
 where  
\(\Lambda_{\tau,\rho,\delta,\cdots} = Z + Z\tau + Z\rho + Z\delta + \cdots \).

In the case of the four arguments define for each \( z , \tau,\rho \in C\) the set of  \( l^7 \)  GTFs with characteristics
 \(\lbrace\cdots,\Theta _{i} (l^{3}z,l^{2}\tau,l^{1}\rho,\delta),\cdots\rbrace\), where 
\( \Theta _{i}~=~\Theta \big [ ^{~~a_{i}~}_{b_{i},2!c_{i},3!d_{i}} \big ] , \hspace{2mm} 0 \leq i \leq l^{7}-1\) , that  with precision 
up to multiplication on some constant represent points of the  projective space  \( P^{ l^{7} -1}_C \), so as in the next paper it will be shown that the roots of GTF with  
different characteristics are different.

For N arguments the dimension of equivariant set of GTF 
will be \( l^{p}\), where \(p = 1 + N(N-1)/2\).  

{\bfseries Statement \ref{sec: GTF with char}.1}

The holomorphic mapping, in the case of the four variables for simplicity, can be defined as:

\( \varphi_{l} :  E_{\tau,\rho,\delta} \to P^{ l^{7} -1}_C , \hspace{4mm} z,\tau,\rho,\delta \to \lbrace\cdots,\Theta _{i} (l^{3}z,l^{2}\tau,l^{1}\rho,\delta),\cdots\rbrace\).

{\bfseries Proof.}

So as \(\lbrace\cdots,\Theta _{i} (z + l,\tau +2! l,\rho + 3! l,\delta+4!l,\cdots),\cdots\rbrace = \lbrace\cdots,\Theta _{i} (z,\tau,\rho,\delta,\cdots),\cdots\rbrace\) and

 \(\lbrace\cdots,\Theta _{i} (z + l\tau + \frac{l^2}{2!} \rho + \frac{l^3}{3!}\delta + \cdots,\hspace{2mm}\tau +  l \rho + \frac{l^2}{2!}\delta  +\cdots
,\hspace{2mm}\rho + l \delta + \cdots,\hspace{2mm}\delta+\cdots,\cdots),\cdots\rbrace =\\
= \lambda  \lbrace\cdots,\Theta _{i} (z,\tau,\rho,\delta,\cdots),\cdots\rbrace\)
where
\( \lambda = exp \lbrace -2\pi i(  lz + \frac{l^2}{2!}\tau + \frac{l^3}{3!} \rho + \frac{l^2}{4!}\delta + \cdots) \rbrace \), then 
explicit action of the extended Heisenberg group \( U_{a;b,2!c,3!d}\) on the embedded set of generalized theta functions with characteristics into projective space \(P^{ l^{7} -1}_C \)is the following:
 
\( \varphi_{l}  ( z + \frac{a}{l}\tau + \frac{1}{2!}(\frac{a}{l})^2\rho + \frac{1}{3!}(\frac{a}{l})^3\delta +\frac{b}{l^3}, \,\,\,
\tau + \frac{a}{l}\rho +\frac{1}{2!}(\frac{a}{l})^2\delta + 2!\frac{c}{l^2},\,\,\,\rho + \frac{a}{l}\delta + 3!\frac{d}{l},\, \,\,\delta) =\\
 = \lbrace\cdots,\Theta _{i} (l^{3}z+\frac{a}{1!}l^{2}\tau +\frac{a^2}{2!}l\rho + \frac{a^3}{3!}\delta +b, \,\,\,l^{2}\tau +\frac{a}{1!}l\rho +\frac{a^2}{2!} \delta + 2!c, 
\,\,\,l^{1}\rho +a\delta + 3!d, \,\,\,\delta),\cdots\rbrace =\\
= \lbrace\cdots, U_{a;b,2!c,3!d} \Theta _{i} (l^{3}z,l^{2}\tau,l^{1}\rho,\delta),\cdots\rbrace = \lbrace\cdots, \sum_{j} c_{ij} \Theta _{j} (l^{3}z,l^{2}\tau,l^{1}\rho,\delta),\cdots\rbrace\), that are linear combinations of \(\Theta _{i} \)-functions according to representation \(U_{a;b,2!c,3!d}\) .


So, generalized theta functions with rational characteristics provide projective embedding of torus 
\( C/\Lambda_{\tau,\rho, \delta,  \cdots}\),  
 where 
\(\Lambda_{\tau,\rho, \delta,  \cdots} = Z + Z\tau + Z\rho + Z\delta + \cdots\).

From the previous construction and group's properties of EHG it evidently follows

{\bfseries Corollary \ref{sec: GTF with char}.2.} The set of \(l^7\)   functions \(\Theta_i \big [ ^{~~~~a_{i}~}_{b_{i},2!c_{i},3!d_{i}} \big ](l^{3}z,l^{2}\tau,l^{1}\rho,\delta)\) for \(N = 4\)  and in the notations  of  section \ref{sec: many parameter}
\(l^p\)   functions \(\Theta_i \big [ ^{~~~~~~~~~~~a_{i}~}_{b_{1},2!b_{2},3!b_{3},...,(N-1)! b_{N-1}}\big ](l^{N-1}\tau_1,l^{N-2}\tau_2,...,l^{1}\tau_{N-1},\tau_N)\) for \(N = 2m, \,\, p = 1 + N(N-1)/2, \)  is equivariant under action of subgroup
\( \Gamma_l = (1,  \,\,la, \,\, l^{N-1}b_1, \,\, 2!l^{N-2}b_2,  \,\, 3!l^{N-3}b_3,..., \,\,(N-1)!l^{1}b_{N-1})\).

{\bfseries Corollary \ref{sec: GTF with char}.3.} There is a chain of tori
\[ P_C^{l^{p(N)}-1}  \begin{array}{c}
                                 \overset{\partial_n^2 \varphi(n)}{\longrightarrow}\\
                                   \underset{\partial_n^{-2} \varphi(n)}{\longleftarrow}
                                 \end{array}  P_C^{l^{p(N-2)}-1}\begin{array}{c}
                                 \overset{\partial_n^2 \varphi(n)}{\longrightarrow}\\
                                   \underset{\partial_n^{-2} \varphi(n)}{\longleftarrow}
                                 \end{array} \cdots\begin{array}{c}
                                 \overset{\partial_n^2 \varphi(n)}{\longrightarrow}\\
                                   \underset{\partial_n^{-2} \varphi(n)}{\longleftarrow}
                                 \end{array}  P_C^{l^{p(4)}-1}  \begin{array}{c}
                                 \overset{\partial_n^2 \varphi(n)}{\longrightarrow}\\
                                   \underset{\partial_n^{-2} \varphi(n)}{\longleftarrow}
                                 \end{array}  P_C^{l^{p(2)}-1} \]
where \( \varphi(n) = \varphi(n;\tau_1,\tau_2,...,\tau_{N-1},\tau_N)\) that determines GTF and  

\(\partial_n^2\varphi(n;\tau_1,\tau_2,...,\tau_{N-1},\tau_N)  =  \varphi(n;\tau_3,\tau_4,...,\tau_{N-1},\tau_N,0,0)\).

\section[Equations for theta functions]{ Equations for generalized  theta functions} 
\label{sec: Equations}

GTF satisfies, besides well known heat equation,
other equations, the simplest of which together with heat equation are: 
 \begin{gather*}
\Theta_t -  \frac{1}{4\pi}\Theta_{zz} = 0, \hspace{20mm}
12\pi\Theta_{\eta} -\Theta_{\tau \tau} = 0 ,  \hspace{20mm}
48\pi^2\Theta_{\delta} + \Theta_{\tau z z} = 0 , \\
4\pi\Theta_{\eta} - \Theta_{\tau \tau} + \Theta_{\rho z} = 0 , 
\hspace{10mm}
\Theta_{zzzz} + 16\pi^2\Theta_{\tau \tau}= 0,\hspace{14mm}
24\pi^2\Theta_{\rho} + \Theta_{zzz} = 0,
\end{gather*}
where \(\tau =it\), \(\delta = i\eta\).

In the first two processes we see that parameter \(\delta\) plays for the parameter \(\tau\) the role of imaginary time so as \(\tau\)~for~\(z\). As~\(t, \eta > 0\), all "times" flow ahead. We have succeeding heat processes from regions of high parameters to the lower one. In these processes parameter \(\tau\) has space character relatively to higher parameter \(\delta\)  and the time character to the lower parameter~\(z\).

This is an example for parameter to have a meaning of "a time" or 'a space' is not absolute but a relative one.

More interesting and not so trivial will be application of GTF for solutions of nonlinear partial differential equations (NPDE). 
Classical TFs with linear form of variables in z argument and arbitrary quadratic form of variables in phase parameter before TF give meromorphic solutions of complete  integrable NPDE using Hirota bi-differential method \cite{Mumford} and  product formula for CTFs.
That is planned to consider in the next papers.

\section[ Theta functions with unified parameters]{ Generalized theta functions with unified view of parameters}
\label{sec: many parameter}

Here we propose unified description of  generalized theta function for any even number \(N = 2m\)  of parameters, denoting them by \(\tau_1 = z, \tau_2 = \tau, ...\) and so on:
\begin{gather}
\Theta\left(\tau_{1},\tau_{2}, ...,\tau_{j}, ...,\tau_N\right)=
\sum_{n\in Z}exp\lbrace2\pi i \sum_{1\le k\le N}\frac{n^k}{k!}\tau_k \rbrace.
\end{gather}

Property of quasi-periodicity has the view:
\begin{gather}
T_a \Theta\left(\tau_{1},\tau_{2}, ...,\tau_{N-1},\tau_N\right) =\nonumber\\
=\Theta\left(\tau_1+\sum_{1\le k\le N-1}\frac{a^{k}}{k!}\tau_{k+1}, \hspace{2mm} \tau_2+\sum_{1\le k\le N-2}\frac{a^k}{k!}\tau_{k+2} , \hspace{2mm}\cdots,\hspace{2mm} \tau_{N-1}+a\tau_N,\hspace{2mm}\tau_N\right)=\nonumber\\
= exp\lbrace -2\pi i \sum_{1\le k\le N}\frac{a^k}{k!}\tau_k \rbrace \sum_{n\in Z+a}exp\lbrace2\pi i \sum_{1\le k\le N}\frac{n^k}{k!}\tau_k \rbrace =\nonumber\\
= exp\lbrace -2\pi i \sum_{1\le k\le N}\frac{a^k}{k!}\tau_k \rbrace \cdot\Theta_a\left(\tau_{1},\tau_{2}, ...,\tau_{N-1},\tau_N\right) .
\end{gather}

Denoting the sum in the exponent by \( \varphi \left(a\right) =  \sum_{1\le k\le N}\frac{a^k}{k!}\tau_k \), for the new arguments of the theta function we have \( \varphi' \left(a\right) =  \tau_1 + \sum_{1\le k\le N-1}\frac{a^k}{k!}\tau_{k+1} \) and
\( \varphi' \left(0\right) = \tau_1 \), \( \varphi^{(m)} \left(a\right) =  \tau_m + \sum_{1\le k\le N-m}\frac{a^k}{k!}\tau_{k+m} \), \( \varphi^{(m)} \left(0\right) =  \tau_m \),
that are connected by the recurrent relation \( \varphi^{(k)} \left(a\right) =  \tau_k + \int\limits_{0}\limits^{a}\varphi^{(k+1)}da \),  the previous fomula takes the form:
\begin{gather}
T_a \Theta\left(\tau_{1},\tau_{2}, ...,\tau_{N-1},\tau_{N}\right) =\
\Theta\left(\varphi' \left(a\right) ,\varphi'' \left(a\right) , \cdots ,\varphi^{(N-1)}\left(a\right),\varphi^{(N)}\right) = \nonumber\\
= exp\lbrace -2\pi i \varphi \left(a\right)  \rbrace \cdot\Theta_a\left(\tau_{1},\tau_{2}, ...,\tau_{N-1},\tau_{N}\right) , 
\hspace{10mm} T_{a1} T_{a2} = T_{a1+a2}.
\end{gather}
 
Action of translation operators \(S\left(b_1,b_2, \cdots, b_N\right)\) and quasi-periodicity operators \(T_a\) are defined as:
\begin{gather}
S\left(b_1,b_2, \cdots,b_N\right)\Theta\left(\tau_1,\tau_2,\cdots,\tau_N\right) = \Theta\left(\tau_1 + b_1,\tau_2 + b_2,\cdots,\tau_N +b_N\right),\\
T_a S\left(b_1,b_2, \cdots,b_N\right)  =  exp\lbrace -2\pi i \sum_{1\le k\le N}\frac{a^k}{k!}b_k \rbrace S\left(b_1,b_2, \cdots,b_N\right)T_a,\\
S\left(b_1,b_2, \cdots, b_N\right)S\left(c_1,c_2, \cdots, c_N\right) =S\left(b_1+c_1,b_2+c_2, \cdots, b_N+c_N\right)   \nonumber
\end{gather}
and form extended Heisenberg group EHG(N) in N=2m dimensions.

\(\Theta\left(\tau_1,\tau_2,\cdots,\tau_N\right) \) function is invariant under translational subgroup \( S\left(b_1,2!b_2, \cdots,k!b_k,\cdots,N!b_N\right)\), where \(b_k\) \(\in Z\).

Evident representation in the space of parameters analogous to the one in section 3 can be marked as \(EH_{N+2}(R)\).

 \section{Conclusions}

Generalization of theta functions on any even number of analytic parameters is suggested that conserve quasi-periodic properties and produce an extension of the Heisenberg group, and their respective non-unitary representations are proposed. 

According to~\cite{Perelomov} coherent states are theta functions and unitary representations of the Heisenberg group 
that is applicable for EHG so as  possible applications of generalized theta functions for solutions of non-linear equations 
by using 
precise values of  roots and factorization of GTF
which are planed to be presented in the next papers so as their projective properties and
many dimensional many parameter generalization of theta functions.

GTF describes succeeding heat processes from regions of high parameters to the lower one. In these processes imaginary parts of parameter \(\tau_{2k}\) 
has a \(\textit{time}\) character relatively to a \(\textit{space}\)  character of a lower parameter \(\tau_k\).

Above it was also shown that generalized theta functions with rational characteristics provide projective embedding of tori 
\( C/\Lambda\),  where  \(\Lambda = Z + Z\tau + Z\rho + Z\delta + \cdots\).



\appendix
\section{Appendix: Proof of statement for GTF}
\label {sec: append}

{\bfseries Statement A.1:} 
{\itshape Finite series \eqref{eq1} with even number of terms \(N =2m\)  is  absolutely and  uniformly convergent in any compact subset \(C^{N-1} \times  H\), which in notations of variables as \(\tau_k\) denoted in section \ref{sec: many parameter} is determined by:
\[|Im( \tau_k)|~<~c_k k!/2\pi, \quad k =1 \div (N-1); \quad Im (\tau_N) > \varepsilon N! /2 \pi\].}
{\bfseries Proof:} Necessary conditions for convergence for \(\forall n\) are: \( |exp \lbrace 2\pi i\sum_k ^N Im (\tau_k)/k! n^k \rbrace| < 1,\)
 that analogously to \cite{Mumford} may be represented as a result of consequent inequalities 
the last of which is 
\[ exp(-\varepsilon n(n-n_1)(n-n_2)...(n-n_{N-1}) < 1,\]
for \( n > n_{N-1} >  n_{N-2} > ... >  n_{1} \), so the series on \(n\) converges and very quickly that is at the same time a sufficient condition.




\pdfbookmark[1]{References}{ref}
\LastPageEnding

\end{document}